\documentstyle[epsfig]{aipproc}

\begin{document}
\title{Angle-Time-Energy Images of\\
Ultra-High Energy Cosmic Ray Sources}
\author{G\"unter Sigl$^*$}
\address{$^*$Department of Astronomy \& Astrophysics\\
Enrico Fermi Institute, The University of Chicago, Chicago, IL
60637-1433}

%\lefthead{LEFT head}
%\righthead{RIGHT head}
\maketitle

\begin{abstract}
Substantial amount of information both on the source and on
characteristics of intercepting magnetic fields is encoded in
the distribution in arrival times, directions, and energies of
charged ultra-high energy cosmic rays from discrete sources. We
present a numerical approach that allows to extract such
information from data from next generation experiments.
\end{abstract}

\section*{Introduction}
The origin of ultra-high energy cosmic rays is still a major
unresolved mystery in astrophysics. It is hard to imagine a
mechanism producing particles of energy up to several
100 EeV ($=10^{20}\,$eV). In addition, sources must be closer than
$\simeq50\,$Mpc because of the limited range of nucleons due to
photo-pion production at these energies. No obvious astrophysical
sources have been found within this
distance~\cite{ssb,elbsom}. Although
deflection of charged primaries can be strong in the direction
of strong magnetic fields along large mass agglomerations such
as the supergalactic plane~\cite{biermann}, a deflection of
several degrees at
the most is expected along most other lines of sight for
nucleons above $10^{20}\,$eV, due to the Faraday rotation limit
on the large-scale field. For next generation experiments with
their anticipated much improved exposure, this opens up the
possibility to do ``particle astronomy'' and pinpoint sources
along the arrival directions.

It has been noted in that respect that a sub class of events
above $4\times10^{19}\,$eV seems to cluster in arrival
directions~\cite{agasa}. If these clusters originated in
discrete sources, some interesting qualitative
consequences result already, such as a limit on the intercepted
magnetic fields that is comparable to the Faraday rotation
limit~\cite{sslh}. Next generation experiments should in this
case see clusters of several tens or even hundreds
of events at these energies in case of the Pierre Auger
Project~\cite{cronin} and the Orbital Wide-angle Light Collector
(OWL)~\cite{owl}, respectively. This
just follows from scaling to the relevant expected exposures.
A data pool of arrival directions, times, and energies of that
size contains a substantial amount of information on both the
source of a given cluster of events and magnetic fields
intercepting the line of sight.

This motivated us to conduct a detailed feasibility study for
the potential of future experiments to reconstruct certain
parameters that characterize the source mechanism and the
large-scale magnetic field both of which are poorly known
at present. We first briefly describe our method and
then summarize our results and give some examples.

\section*{Description of Approach}
Here we describe the essential ingredients of our numerical
approach; more details can be found in Refs.~\cite{slo,sl}.

The propagation of nucleons through extragalactic space is
simulated using the Monte Carlo technique: First, a magnetic
field realization is set up on a grid via Fast Fourier
Transformation by sampling a power spectrum of the form
$\left\langle B^2(k)\right\rangle\propto k^{n_B}$ for wavenumbers
$k<2\pi/l_c$ and 0 otherwise, where $l_c$ characterizes the
coherence scale and $n_B$ the magnetic field power
spectrum. As results are quite insensitive to $n_B$, we assume
$n_B=0$ if not stated otherwise. More sophisticated models for
the magnetic field
including the role of the large scale structure of galaxies, for
example, along the lines discussed in Ref.~\cite{biermann}, may
be implemented in the future. Many nucleon trajectories are then
calculated between a given source and observer by sampling
direction of emission, injection energy and the stochastic pion
production loss that becomes important above the
Greisen-Zatsepin-Kuzmin (GZK) cut-off~\cite{gzk} at a few
$10^{19}\,$eV . Pair production by protons has been incorporated
as a continuous energy loss. One of the
main problems that has to be solved when images
of discrete sources are discussed, has not been considered in
other work on propagation and deflection~\cite{codes} and consists of
the fact that different trajectories not only originate at the
same source, but also have to reach the same observer.

From the injection energies, direction, time, and energy of
arrival recorded for the trajectories we then calculate
histograms for the distribution in these quantities by
convolving with the injection spectrum (typically a power law
with index $\gamma$ for the differential spectrum) in energy and
with a timescale $T_{\rm S}$ that characterizes the emission
timescale. Histograms are also smeared out in energy to account
for finite energy resolution (typically $\Delta E/E\simeq0.14$,
a value expected for future detectors) and are proportional to the
source fluence $N_0$. We also use the parameter
\begin{equation}
  \tau_E\,\simeq\,2.0\,
  \left(\frac{D}{30\,{\rm Mpc}}\right)^2
  \left(\frac{E}{100\,{\rm EeV}}\right)^{-2}
  \left(\frac{B_{\rm rms}}{10^{-11}\,{\rm G}}\right)^2
  \left(\frac{l_{\rm c}}{1\,{\rm Mpc}}\right)\;{\rm yr}.
  \label{t_delay}
\end{equation}
which is the average time delay for a proton of energy $E$ over a
distance $D$ in a field of r.m.s. strength $B_{\rm rms}$ when
energy loss is negligible, and $\tau_E\ll D$~\cite{wm}. It is
related to the average deflection angle $\theta_E$ by
\begin{equation}
  \theta_E\simeq0.02^{\circ}\left(\frac{D}{10\,{\rm Mpc}}\right)^{-1/2}
  \left(\frac{\tau_E}{1\,{\rm yr}}\right)^{1/2}\,.\label{ang}
\end{equation}
The subscript $E$ is given in EeV in the following.

Clusters of events are then obtained by sampling the histogram
with Poisson statistics over a time window of width $T_{\rm
obs}$ which constitutes the experimental lifetime, at a random
position. Conversely, for a given cluster of events, a
likelihood can be calculated for a given histogram that
corresponds to certain values of the physical parameters
described above. Averaging over different observational window
positions and different realizations of the magnetic field for
the same parameters yields the likelihood function ${\cal
L}(\tau_{100},T_{\rm S},D,\gamma,N_0,l_c,n_B)$. Marginalization
over part of these parameters, using priors that account for
certain constraints and other available information, can be used
to reduce the parameter space.

\section*{Results and Examples}
We first give a brief outline of the main features of the
angle-time-energy images of clusters of ultra-high energy
nucleons which have been described in detail in Ref.~\cite{lsos}.

If both $T_{\rm S}<\tau_{100}$, and $\tau_{100}$ is small
compared to $T_{\rm obs}$, arrival time and energy are
correlated according to $\tau_E\propto E^{-2}$; see
Eq.~(\ref{t_delay}). The angular image can not be resolved in
this case.

\begin{figure}[b!]
\centerline{\epsfig{file=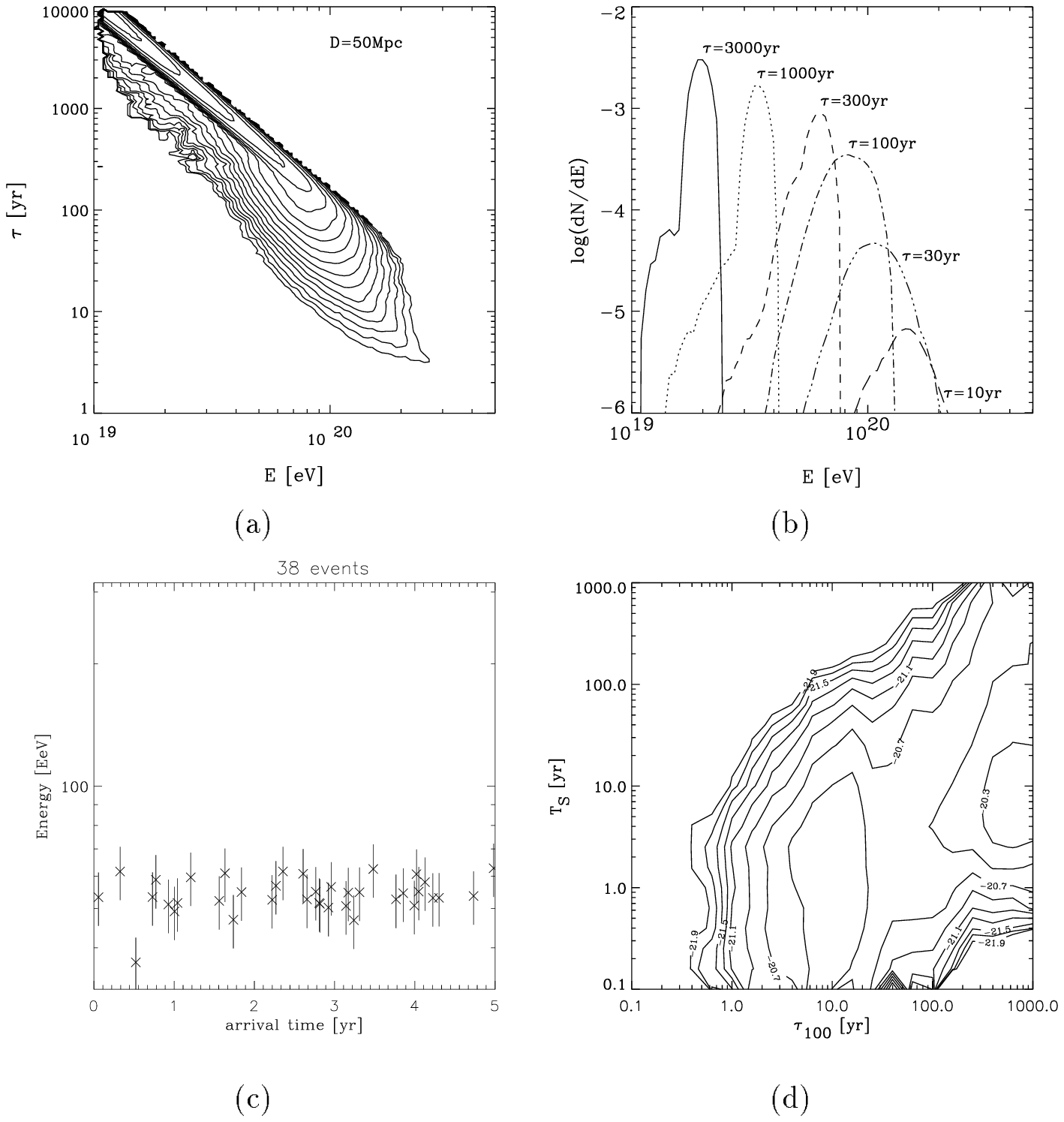,height=6in,width=6in}}
\vspace{10pt}
\caption[...]{(a) An arrival time-energy histogram for
$\gamma=2.0$, $\tau_{100}=100\,$yr, $T_{\rm S}\ll\tau_{100}$,
$l_c\simeq1\,$Mpc, $D=50\,$Mpc, corresponding to $B_{\rm
rms}\simeq4\times10^{-11}\,$G. Contours are in steps of a factor
$10^{0.4}=2.51$; (b) Observable
energy spectrum for several positions of the observational
window in the histogram in (a); (c) Example for a cluster in the
arrival time-energy plane resulting from one of the cuts shown
in (b); (d) The likelihood function, marginalized over $N_0$
and $\gamma$, for $D=50\,$Mpc, $l_c=0.25\,$Mpc, for the cluster
shown in (c), in the $T_{\rm S}-\tau_{100}$
plane. The contours shown go from the maximum down to about 0.01
of the maximum in steps of a factor $10^{0.2}=1.58$. The
fall-off at $\tau_{100}\gtrsim50\,$yr and $T_{\rm
S}\lesssim3\,$yr is a numerical artifact due to limited
statistics. The true parameters are reasonably well
reconstructed.}
\label{F1}
\end{figure}

A source, such that 
$\tau_{100}\gg T_{\rm S}$ and $\tau_{100}\gg T_{\rm obs}$, can 
be seen only in a limited range of energies, at a given time, as
first pointed out in Ref.~\cite{wm}, and demonstrated in
Fig.~\ref{F1}. Below the GZK cut-off, the width of
this stripe, in the time-energy plane and within the
observational window of length $T_{\rm obs}$, is then governed
by the ratio $D\theta_E/l_c$ for the energy at which events are
observed: If this ratio is much smaller than 1,
all nucleons have experienced the same magnetic field 
structure during their propagation and the width is 
very small in the absence of pion production; in the opposite
case the width is expected to be $\Delta\tau_E/\tau_E\sim60$\%,
even for negligible energy loss. Furthermore, the angular image
is point-like or diffuse, with $\theta_E$ describing the
systematic off-set from the direction to the source, or the
angular extent of the diffuse image that is centered on the
source, respectively, in these two cases.
If $D\theta_E/l_c\sim1$, several images of the
source can result~\cite{progress}.

\begin{figure}[b!]
\centerline{\epsfig{file=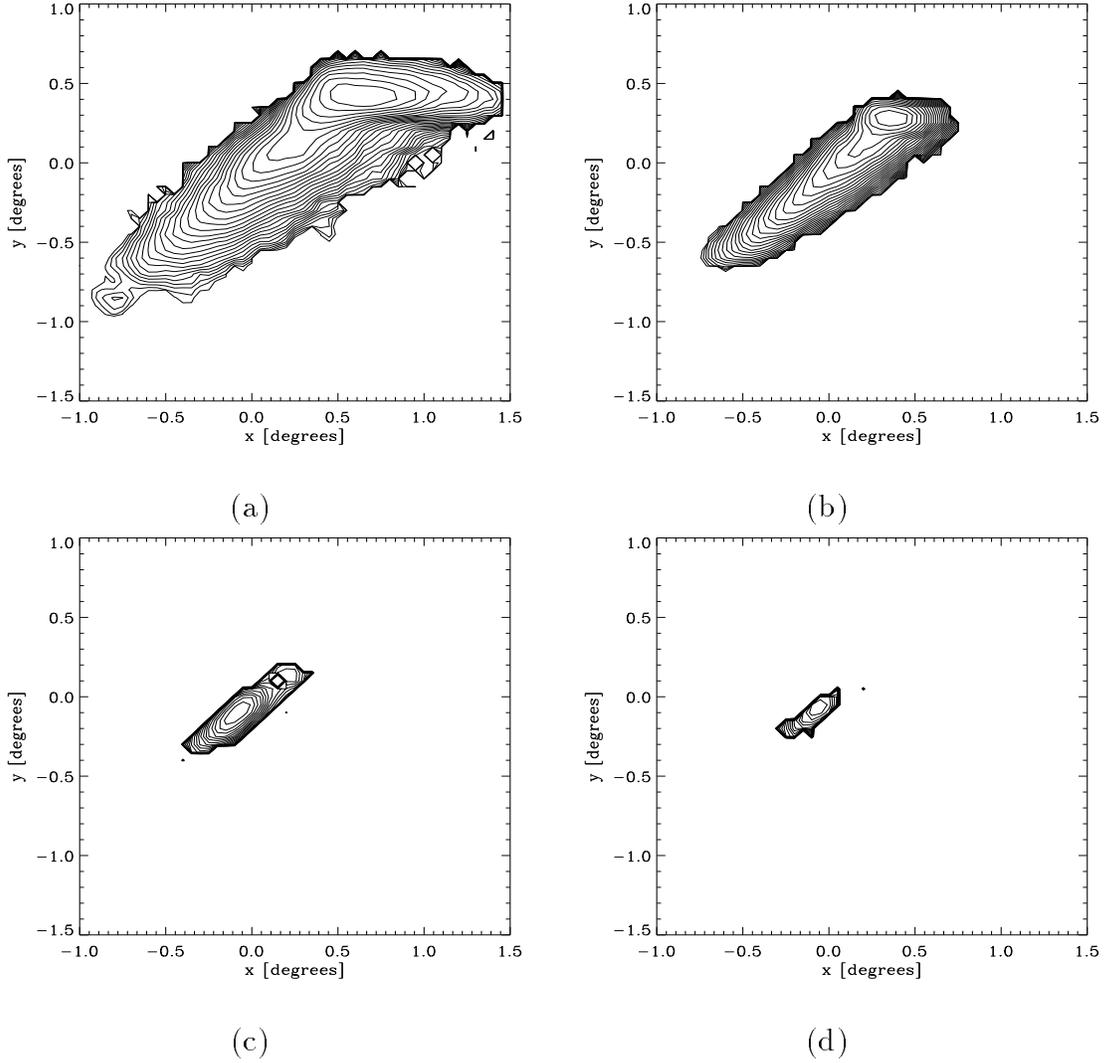,height=6in,width=6in}}
\vspace{10pt}
\caption[...]{An angle-histogram for $\gamma=2.0$,
$\tau_{100}=10^4\,$yr, $T_{\rm S}\gg\tau_{100}$,
$l_c\simeq1\,$Mpc, $D=50\,$Mpc,
corresponding to $B_{\rm rms}\simeq4\times10^{-10}\,$G.
An angular resolution of 0.05$^\circ$ was assumed.
The point $x=y=0$ is the source
position and the contours decrease in steps of 0.1 in the
logarithm to base 10.
(a) Image integrated over all energies $E>30\,$EeV. Two
partially blended, elongated images at
$x\simeq0.2^\circ$, $y\simeq0.2^\circ$ and at
$x\simeq0.7^\circ$, $y\simeq0.4^\circ$ are clearly visible, the
second one being more luminous by about a factor 4; (b) Same for
$E>100\,$EeV. The two images are now closer to
the source position; (c) Same for $E>200\,$EeV. The second image
has almost disappeared; (d) Same for $E>300\,$EeV. As a consequence,
$D\theta_E/l_c\simeq1$ for $E\simeq100$EeV. If $D$ can
be estimated from the energy spectrum, an estimate for $l_c$
results.}
\label{F2}
\end{figure}

For a source emitting continuously at all
energies of interest here, {\it i.e.} with $T_{\rm S}\gg\tau_{30}$ and
$T_{\rm S}\gg T_{\rm obs}$, events of any energy can be recorded
at any time. Whereas the above remarks on the angular image
now apply for all energies (see Fig.~\ref{F2}), the distribution
of arrival time {\it vs.} energy is now uniform.

\begin{figure}[b!]
\centerline{\epsfig{file=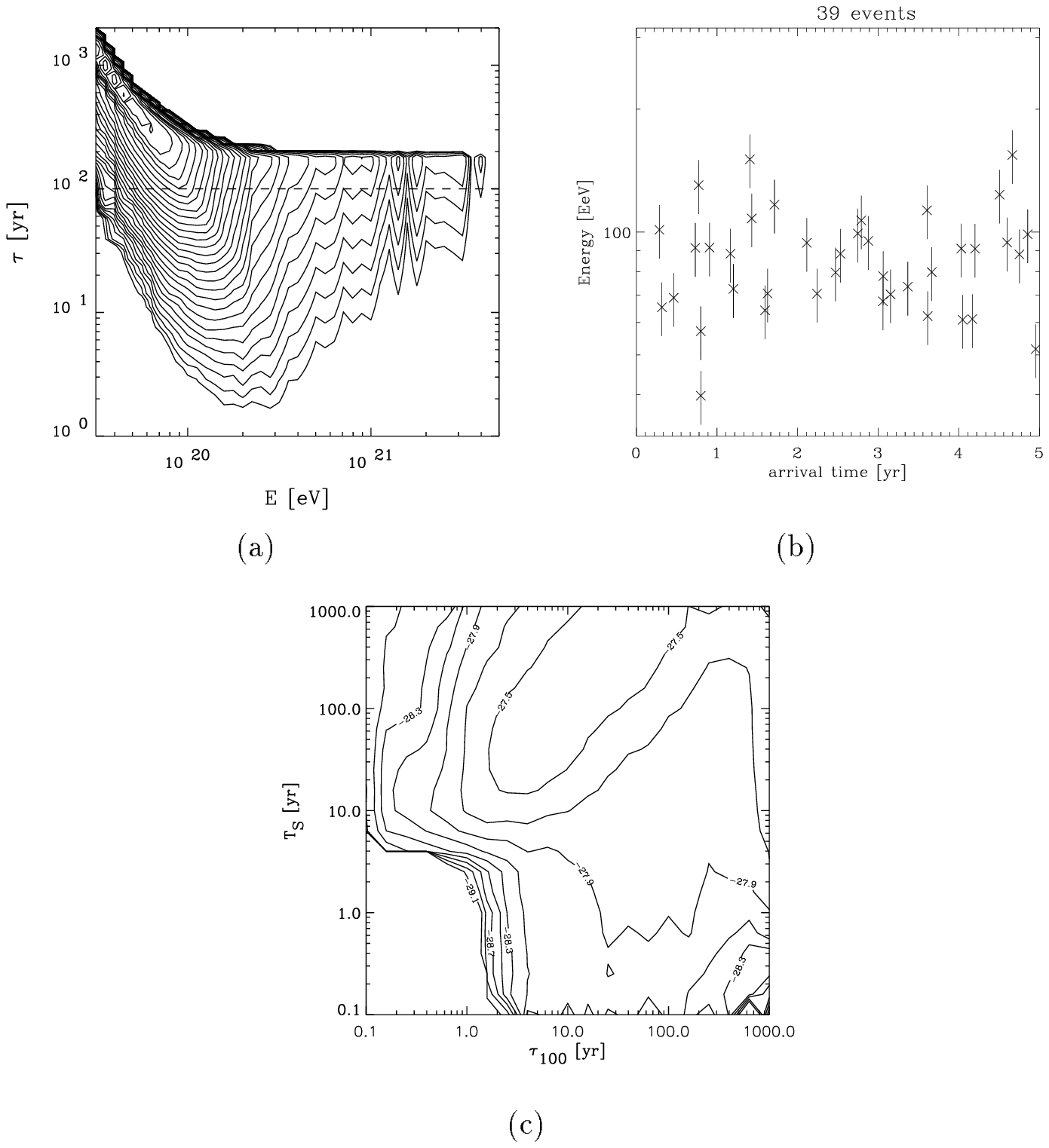,height=6in,width=6in}}
\vspace{10pt}
\caption[...]{(a) An arrival time-energy histogram for
$\gamma=2.0$, $\tau_{100}=50\,$yr, $T_{\rm S}=200\,$yr,
$l_c\simeq1\,$Mpc, $D=50\,$Mpc, corresponding to $B_{\rm
rms}\simeq3\times10^{-11}\,$G. Contours are in steps of a factor
$10^{0.4}=2.51$;
(b) Example for a cluster in the
arrival time-energy plane resulting from the cut indicated in
(a) by the dashed line at $\tau\simeq100\,$yr; (c) Same as
Fig.~\ref{F1} (d), but for the cluster shown in (b). Note that
the likelihood clearly favors $T_{\rm S}\simeq\tau_{50}$. For
$\tau_{100}$ large enough to be estimated from the angular
image size, $T_{\rm S}\gg T_{\rm obs}$ can, therefore, be
estimated as well.}
\label{F3}
\end{figure}

Finally, for a source, such that $\tau_{100}< T_{\rm S}$ and
$\tau_{30}> T_{\rm S}$, together with $T_{\rm S}\gg T_{\rm obs}$,
there exists an energy $E_{\rm C}$, such 
that $\tau_{E_{\rm C}}=T_{\rm S}$. In this case, protons with
$E<E_{\rm C}$ are not detected, as they could 
not have reached us within $T_{\rm obs}$. However, protons with
$E>E_{\rm C}$ are detected as for a continuously emitting
source, {\it i.e.} with a uniform distribution of arrival times
{\it vs.} energy (see, e.g., Fig.~\ref{F3}).

We now summarize results on the potential to reconstruct the
parameters $\tau_{100}$, $T_{\rm S}$, $D$, $\gamma$, $N_0$,
$l_c$, and $n_B$ in these scenarios. Details have been presented
in Ref.~\cite{sl}.

The likelihood presents different degeneracies between
different parameters, which complicates the analysis. As an example,
the likelihood is degenerate in the ratios $N_0/T_{\rm S}$, or
$N_0/\Delta\tau_{100}$, with $N_0$ the total fluence, and
$\Delta\tau_{100}$ the spread in arrival time: these ratios
represent rates of detection. Another example is given by the
degeneracy between the distance $D$ and the injection energy 
spectrum index $\gamma$. Yet another is the ratio
$D\theta_E/l_c\propto(D\tau_E)^{1/2}/l_c$,
that controls the size of the scatter around the mean of the 
$\tau_E-E$ correlation. Therefore, in most general cases, values for 
the different parameters cannot be pinned down, and generally, only 
domains of validity are found. We remark, however, that the
generic scenarios discussed above are, in general, easy to
distinguish from the likelihood function (see, e.g.,
Fig.~\ref{F2}).

We find that the distance to the source is obtained from the pion 
production signature, above the GZK cut-off, when 
the emission timescale of the source dominates over the time delay. 
The lower the minimal energy above which the source appears as
emitting continuously, the higher the accuracy on the distance
$D$. The error on $D$ is, in the 
best case, typically a factor 2, for one cluster of $\simeq40$ events.
In this case, where the emission timescale dominates over the time 
delay at all observable energies,
information on the magnetic field is only contained in the
angular image. Qualitatively, the size of the angular image is 
proportional to $B_{\rm rms}(Dl_c)^{1/2}/E$, whereas the structure
of the image, {\it i.e.} the number of separate images, is
controlled by the ratio $D\theta_E/l_c\propto D^{3/2}B_{\rm
rms}/El_c^{1/2}$. Finally,
the case where the time delay dominates over the emission
timescale, with a time delay shorter than the lifetime of the
experiment, also allows to estimate the distance 
with a reasonable accuracy. 

The injection spectrum index $\gamma$ can be measured provided 
ultra-high energy cosmic rays are recorded over a bandpass in
energy that is sufficiently broad. In general,
it is comparably easy to rule out a hard injection
spectrum if the actual $\gamma\gtrsim2.0$, but it is much harder
to distinguish between $\gamma=2.0$ and 2.5.

The strength of the magnetic field can only be obtained from the 
time-energy image in this latter case because the angular image
will not be resolvable. When the time delay 
dominates over the emission timescale, and is, at the same time,
larger than the lifetime $T_{\rm obs}$ of the experiment, only a
lower limit corresponding to $T_{\rm obs}$, can be placed on the time 
delay, hence on the strength of the magnetic field. When combined with 
the Faraday rotation upper limit, this would nonetheless allow
to bracket the r.m.s. magnetic field strength within a
few orders of magnitude. Here as well, significant information is
contained in the angular image.

The coherence length enters the ratio $(D\tau_E)^{1/2}/l_c$ that 
controls the scatter around the mean of the $\tau_E-E$ correlation in 
the time-energy image. It can therefore be estimated from the width of 
this image, provided the emission timescale is dominated by $\tau_E$ 
(otherwise the correlation would not be seen), and some prior 
information on $D$ and $\tau_E$ is available. If the source
appears continuous and the time delay is large enough to resolve
the angular image, $l_c$ can be constrained or even estimated
from the fact that $D\theta_E/l_c$ passes through 1 at the
energy where the scatter $\Delta\theta_E/\theta_E$ becomes
comparable to 1 (it is much smaller at energies that are higher
but still below the GZK cut-off; see Fig.~\ref{F2}). Our
simulations showed no sensitivity to the magnetic field power
spectrum characterized by $n_B$.

An emission timescale much larger than the experimental lifetime
may be estimated if a lower cut-off in the spectrum is
observable at an energy $E_{\rm C}$, indicating that $T_{\rm
S}\simeq\tau_{E_{\rm C}}$. The latter may, in turn, be estimated
from the angular image size via Eq.~(\ref{ang}), where the
distance can be estimated from the spectrum visible above the
GZK cut-off, as discussed above. An example
for this scenario is shown in Fig.~\ref{F3}. For angular
resolutions $\Delta\theta$, timescales in the range
\begin{equation}
  3\times10^3\,\left(\frac{\Delta\theta}{1^\circ}\right)^2
  \left(\frac{D}{10\,{\rm Mpc}}\right)\,{\rm yr}
  \lesssim T_{\rm S}\simeq\tau_E\lesssim10^4\cdots10^7\,
  \left(\frac{E}{100\,{\rm EeV}}\right)^{-2}\,{\rm yr}
  \label{tsscale}
\end{equation}
could be probed. The lower limit follows from the requirement
that it should be possible to estimate $\tau_E$ from $\theta_E$,
using Eq.~(\ref{ang}), otherwise only an upper limit on $T_{\rm
S}$, corresponding to this same number, would apply.
The upper bound in Eq.~(\ref{tsscale}) comes from constraints on
maximal time delays in cosmic magnetic fields, such as the Faraday
rotation limit in case of the cosmological large-scale field
(smaller number) and knowledge on stronger fields associated
with the large-scale galaxy structure (larger
number). Eq.~(\ref{tsscale}) constitutes an interesting range of
emission timescales for many conceivable scenarios of ultra-high
energy cosmic rays. For example, the hot spots in certain
powerful radio galaxies that have been suggested as ultra-high
energy cosmic ray sources~\cite{rb}, have a size of only several
kpc and could have an episodic activity on timescales of
$\sim10^6\,$yr.

\section*{Conclusions}
A wealth of information on both the production mechanism of
highest energy cosmic rays and on the structure of large-scale
magnetic fields is encoded in angle-time-energy images of
discrete sources. If the clustering suggested by AGASA
is real, tens (for the Pierre Auger Project) to hundreds (for
the OWL Project) of events above a few $10^{19}\,$eV
can be expected from individual sources alone. With
resolutions of 10-20\% in energy and fractions of a
degree in angle, next generation experiments should be able to
exploit this information.

Special thanks go to Martin Lemoine for an ongoing extensive
collaboration on this subject. I also thank Angela Olinto and
David Schramm for collaboration in earlier stages. This work was
supported, in part, by the DoE, NSF, and NASA at the University
of Chicago.

\begin{references}

\bibitem{ssb} G.~Sigl, D.~N.~Schramm, and P.~Bhattacharjee,
{\it Astropart.~Phys.} {\bf 2}, 401 (1994).

\bibitem{elbsom} J.~W.~Elbert, and P.~Sommers, {\it
Astrophys.~J.} {\bf 441}, 151 (1995).

\bibitem{biermann} for a discussion of this case see, e.g.,
P.~L.~Biermann, H.~Kang, J.~P.~Rachen, and D.~Ryu, e-print
astro-ph/9709252, to appear in Proc. Moriond
Meeting on High Energy Phenomena, Jan. 1997, Les Arcs.

\bibitem{agasa} N.~Hayashida et al., {\it Phys.~Rev.~Lett.} {\bf
77}, 1000 (1996).

\bibitem{sslh} G.~Sigl, D.~N.~Schramm, S.~Lee, and C.~T.~Hill,
{\it Proc.~Natl.~Acad.~Sci.~USA} {\bf 94}, 10501 (1997).

\bibitem{cronin} J.~W.~Cronin, {\it Nucl. Phys. B (Proc. Suppl.)}
{\bf 28B}, 213 (1992); The Pierre Auger Observatory Design Report
(2nd ed.) 14 March 1997.

\bibitem{proc} Proc. of
{\it International Symposium on Extremely High Energy Cosmic Rays:
Astrophysics and Future Observatories} (Institute for Cosmic Ray
Research, Tokyo, 1996).

\bibitem{owl} J.~F.~Ormes et al., in {\it Proc.~25th International
Cosmic Ray Conference} (Durban, 1997), eds.: M.~S.~Potgieter et
al., 5, 273; Y.~Takahashi et al., in~\cite{proc}, p.~310.

\bibitem{slo} G.~Sigl, M.~Lemoine, and A.V.~Olinto, {\it
Phys.~Rev.~D} {\bf 56}, 4470 (1997).

\bibitem{sl} G.~Sigl and M.~Lemoine, e-print astro-ph/9711060,
submitted to {\it Astropart.~Phys.}

\bibitem{gzk} K.~Greisen, {\it Phys.~Rev.~Lett.} {\bf 16}, 748
(1966); G.~T.~Zatsepin and V.~A.~Kuzmin, {\it Pis'ma
Zh. Eksp. Teor. Fiz.} {\bf 4}, 114 (1966) [{\it JETP. Lett.}
{\bf 4}, 78 (1966)].

\bibitem{codes} see, e.g., G.~A.~Medina Tanco, E.~M.~de Gouveia
Dal Pino, and J.~E.~Horvath, {\it Astropart.~Phys.} {\bf 6}, 337
(1997); R.~Lampard, R.~W.~Clay, and B.~R.~Dawson, {\it
Astropart.~Phys} {\bf 7}, 213 (1997).

\bibitem{wm} E.~Waxman and J.~Miralda-Escud\'{e}, {\it
Astrophys.~J.} {\bf 472}, L89 (1996).

\bibitem{lsos} M.~Lemoine, G.~Sigl, A.~V.~Olinto, and D.N. 
Schramm, {\it Astrophys.~J.} {\bf 486}, L115 (1997).

\bibitem{progress} M.~Lemoine and G.~Sigl, work in progress.

\bibitem{rb} J.~P.~Rachen, and P.~L.~Biermann, {\it
Astron.~Astrophys.} {\bf 272}, 161 (1993).

\end{references}
\end{document}